# Wafer-Scale Integration of Piezo- and Ferroelectric $Al_{0.64}Sc_{0.36}N$ Thin Films by Reactive Sputtering

Sanjay Nayak*[1], Venkata Raveendra Nallagatla[1], Ravindra Singh Bisht[1], Dmytro Solonenko[1], Demian Henzen[2], Washim Reza Ali[1], Carla Maria Lazzari[3], Robert JW Frost[4], Andrea Serafini[3], Davide Codegoni[3], Amalia Balsamo[5], Rossana Scaldaferri[5], Elaheh Allahyari[5], Anirban Ghosh[1], Martin Kratzer[2], Andrea Picco[3], Sonia Costantini[3], Andrea Rusconi[6], Mohssen Moridi[1], Humberto Campanella[1], Marco Deluca[1], and Annalisa De Pastina[1]

[1] Silicon Austria Labs GmbH, Villach 9524, Austria.
[2] Evatec AG, Trübbach-9477, Switzerland.
[3] STMicroelectronics, Agrate Brianza, 20864, Italy.
[4] Tandem Laboratory, Uppsala University, Uppsala, SE 75120, Sweden.
[5] STMicroelectronics, 80022 Arzano, Naples, Italy.
[6] USound GmbH, Graz-8020, Austria.

*Corresponding Author: sanjay.nayak@silicon-austria.com, sanjaynayak.physu@gmail.com (SN)

## Abstract

Large-area deposition of Aluminium-Scandium-Nitride ($Al_{1-x}Sc_xN$) thin films with higher Sc content (x) remains challenging due to issues such as abnormal orientation growth, stress control, and the undesired crystal phase. These anomalies across the wafer hinder the development of high scandium-content AlScN films, which are critical for microelectromechanical systems applications. In this study, we report the sputter deposition of $Al_{0.64}Sc_{0.36}N$ thin films from a 300 mm $Al_{0.64}Sc_{0.36}$ alloy target on 200 mm Si (100) wafers, achieving an exceptionally high deposition rate of 8.7 μm/h with less than 1% AOGs and controllable stress tuning. Comprehensive microstructural and electrical characterizations confirm the superior growth of high-quality $Al_{0.64}Sc_{0.36}N$ films with exceptional wafer-average piezoelectric coefficients ($d_{33,f}$=15.62 pm/V and $e_{31,f}$= -2.9 C/m$^2$) owing to low point defects density and grain mosaicity. This was accomplished through the implementation of an optimized seed layer and a refined electrode integration strategy, along with optimal process conditions. The wafer yield and device failure rates are analysed and correlated with the average stress of the films and their stress profiles along the diameter. The resulting films show excellent uniformity in structural, compositional, and piezoelectric properties across the entire 200 mm wafer, underscoring their strong potential for next-generation MEMS applications.

## I. INTRODUCTION

Green and sustainable microelectronics aim to minimize energy consumption, reduce reliance on scarce or toxic materials, and integrate seamlessly with existing manufacturing processes to lower environmental impact[1]. Central to this endeavour is the development of advanced materials that enhance device efficiency while supporting scalability and eco-friendliness. Aluminium Nitride (AlN) is one of a well-established wide bandgap III-V sustainable semiconductors with a wurtzite phase crystal structure, widely utilized in microelectromechanical systems (MEMS) due to its intrinsic piezoelectric properties, high thermal conductivity, high thermo-chemical stability, and its manufacturing compatibility with complementary metal-oxide-semiconductor (CMOS) technology[2]. These properties make AlN ideal for applications such as acoustic wave filters[3–5], sensors[6–11], and transducers[12–15]. However, un-doped AlN has limitations, including a relatively modest effective electromechanical coupling factor, $k_{eff}^2 \approx 5.2$-$5.8\%)$[16,17] and lack of ferroelectricity due to its inability to switch polarization under an electric field below its dielectric breakdown limit [18], which restrict its performance in high-frequency and reconfigurable devices, crucial components in creating sustainable and adaptable systems for future electronic and optical devices.

In 2009, Akiyama et al. discovered that alloying of AlN with Scandium Nitride (ScN) resulted into the formation of wurtzite Aluminum-Scandium-Nitride ($Al_{1-x}Sc_xN$) exhibiting anomalous piezoelectric behaviour, with a 400% increment in effective piezoelectric response ($d_{33,f}$) with 43% of ScN in AlScN alloy thin films[19,20], leading to an effective $k_{eff}^2$ increase by 2.6 times compared to AlN in bulk acoustic wave (BAW) resonators[17]. Moreover, $Al_{1-x}Sc_xN$ also exhibits ferroelectricity. Studies have shown that $Al_{1-x}Sc_xN$ with Sc content ($x \sim 0.3$) demonstrates a large remanent polarization exceeding 100 μC/cm² and a coercive field ($E_c$) of approximately 3 MV/cm[21]. Ferroelectricity in AlScN is enabled by Sc dopant induced structural distortion in the wurtzite lattice that lowers the energy barrier for polarization switching. Ferroelectricity in $Al_{1-x}Sc_xN$ opened a plethora of reconfigurable device applications such as ferroelectric field-effect transistor (FeFET)[22], compute-in-memory[23], and electro-optic duplex memristor[24].

Despite its promising initial success, the widespread adoption of $Al_{1-x}Sc_xN$ faces several materials and integration challenges. These include difficulties in stabilizing its wurtzite phase with high Sc content[25,26], the pervasive formation of abnormally oriented grains (AOGs)[27], high oxidation sensitivity[28], high polarization switching voltages[18,21], and high leakage currents[29]. Beyond these material-specific hurdles, a major technological barrier to commercialization lies in developing industrially relevant thin films that exhibit uniform microstructure and consistent electromechanical properties across large-scale wafers, in particular for wafer scale thin film depositions on 200 mm diameter silicon substrates.

Various methods have been employed for the thin-film growth of $Al_{1-x}Sc_xN$, including molecular beam epitaxy (MBE)[30–32], metal-organic chemical vapor deposition (MOCVD)[33,34], and reactive magnetron sputtering[35,36]. While MBE and MOCVD are known for producing exceptionally high-quality thin films, their commercial viability is often limited by high costs, film deposition rates, and scalability challenges. Furthermore, these methods often employ processes that are not aligned with the United Nations' "17 Sustainable Development Goals"[37]. In contrast, for the microelectronics industry, especially in MEMS where AlN serves as a functional thin film, magnetron sputtering has proven to be an effective alternative. As a physical vapor deposition technique, magnetron sputtering utilizes solid metal targets with pure Ar and $N_2$ gases, ensuring that no harmful byproducts are generated. This method also boasts lower production costs, inherent scalability, and is a well-established technique within the industry.

Work of Akiyama et al.[19,20] shows a consistent increase in $d_{33,f}$ of $Al_{1-x}Sc_xN$ with Sc content ($x$) increased up to 0.43. The increase in $x$ also lowers the $E_C$ but also the saturation polarization[21]. The mechanical properties of $Al_{1-x}Sc_xN$ films are also influenced by the fraction of ScN content in it. For example, with $x > 0.25$ the material's elastic properties change significantly, with the Young's modulus decreasing from 303.0 GPa (undoped AlN) to 217.1 GPa ($Al_{0.75}Sc_{0.25}N$), and the Poisson's ratio increasing from 0.236 to 0.348, indicating elastic softening and increased elastic deformation[38,39]. Increase of $x$ in $Al_{1-x}Sc_xN$ also risks increasing the fraction of phase transition from wurtzite to cubic and leads to formation of high densities of AOGs[27] that increases surface roughness and consequently results into high electrical leakage current. There is a consensus in the research community now that up until $x \approx 0.30$, good quality $Al_{1-x}Sc_xN$ thin films can be deposited and indeed they are being done routinely[40–45]. However, beyond $x > 0.30$, the implementation challenges persist.

While co-sputtering, which involves the separate sputtering of Al and Sc metal targets with finely tuned power to control film composition[21,35,46–48], has proven effective for a lower dimensional substrate its suitability for high-throughput industrial production of $Al_{1-x}Sc_xN$ is limited. The co-sputtering method faces significant challenges in maintaining consistent film composition over time and throughout the target's lifetime[49]. Additionally, co-sputtering typically yields a low deposition rate, and achieving uniform film thickness and stress distribution within acceptable tolerance limits remains problematic. Furthermore, the ability to tune stress and wafer bow without compromising the crucial film orientation and crystallinity is often lacking. For industrial applications, the preferred approach is to utilize a single, large alloy target (e.g., 300 mm diameter $Al_{1-x}Sc_x$) to deposit films uniformly onto large-scale substrates, such as 200 mm diameter silicon wafers. This single-target approach offers superior consistency, higher deposition rates, and improved control over film properties essential for high-volume manufacturing.

Barth et al.[50] used an $Al_{0.70}Sc_{0.30}$ alloyed target and sputtered c-axis textured AlScN thin films on 200 mm Si wafers. The full width at half maximum of x-ray rocking curve (XRC-FWHM) of AlScN's 0002 peak was $> 2.2°$, while a broad range of AFM surface roughness ($S_q \approx 1.4$-9.5 nm) was observed across the wafer surface. By using a specialized magnetron target design, where Sc pellets were embedded within an Al plate, Österlund et al.[51] deposited $Al_{0.70}Sc_{0.30}N$ films on a 150 mm substrate, achieving an XRC-FWHM of 3.36° with a surface roughness, $S_q$ of 1 nm. Sputtering from a 101.6 mm diameter $Al_{0.70}Sc_{0.30}$ alloy target on 200 mm Si wafer, Nie et al.[52] demonstrated films with an XRC-FWHM of 2.5°. Similarly, work by Pirro et al. [53] shows an XRC-FWHM of 2.4°. Very recently, Liffredo et al.[54] used a 200 mm alloyed target of $Al_{0.60}Sc_{0.40}$ and deposited film on a 100 mm Si wafer, resulting in an XRC-FWHM of 1.6°. The $d_{33}$ and $e_{31}$ values were reported to be 29±3 pC/N and -0.85 C/m², respectively. Kusano et al.[55] deposited $Al_{0.64}Sc_{0.36}N$ thin films on a 150 mm Silicon-on-Insulator (SOI) wafer, where the XRC-FWHM varied from 1.5° (center) to 2.3° (edge). The effective transverse piezoelectric stress coefficient, $e_{31,f}$, was estimated to be around −1.58 C/m². Using a 101.6 mm $Al_{0.64}Sc_{0.36}$ alloyed target and films deposited on a 200 mm SOI substrate, Fichtner et al. [21] demonstrated an $e_{31,f}$ of -2.90 C/m². In this work, it was not explicit whether the mentioned value was from the center of the wafer or an average value from the wafer. Very recently, Kreutzer et al. [56] reported sputtered $Al_{0.60}Sc_{0.40}N$ thin films from a 300 mm alloyed target onto a 200 mm SOI substrate with a high deposition rate of 3.3 μm/h utilizing an industrial tool with a wide range of XRC-FWHM varied from 1.4° (from center) to 5.01° (at the edge). The wafer average $d_{33,f}$, dielectric constant ε, and tangent loss functions (tan δ) were 15.62 pm/V, 28.32, and 0.57%, respectively. While individual studies have demonstrated the feasibility of depositing AlScN thin films at the wafer scale and highlighted improvements in Sc content and piezoelectric properties through various process parameters, a comprehensive wafer-scale mapping of structural, mechanical, and piezoelectric properties is yet to be benchmarked for real industrial scale device production with a high Sc content film ($x > 0.30$).

In this work, we report on the sputtered growth of $Al_{0.64}Sc_{0.36}N$ thin films on 200 mm Si (100) wafers, utilizing a 300 mm $Al_{0.64}Sc_{0.36}$ alloy target. This particular Sc composition was strategically chosen due to its proximity to the phase boundary[55], the commercial availability of the target, and to mitigate the risk of target damage and brittleness associated with intermetallic phases[57,58]. We report on the sputtered growth of $Al_{0.64}Sc_{0.36}N$ thin films on 200 mm Si (100) wafers with a high deposition rate. Through comprehensive microstructural and electrical characterization, we demonstrate the successful growth of exceptionally high-quality $Al_{0.64}Sc_{0.36}N$ thin films on Si substrates. This was achieved by employing an appropriate seed layer and optimizing the electrodes adoption strategy. The resulting thin films exhibit outstanding structural, compositional, and electrical property uniformity across the entire 200 mm wafer showcasing promises for future all nitride electronics. We further emphasize the critical role of microstructure and residual stress in influencing the electrical properties and overall wafer yield of these AlScN thin films.

## II. METHODS

AlScN thin films were deposited via pulsed direct current reactive sputtering, utilizing a single target module of the CLUSTERLINE 200E (Evatec AG) tool. The sputter chamber incorporated a circular 300 mm $Al_{0.64}Sc_{0.36}$ target with a purity of 99.9wt% (JX Advanced Metals Corporation) and a gas flow rate ratio of Ar to $N_2$ was maintained at 1:2 throughout the deposition process. Deposition was performed on 200 mm Si (100) wafers, with the substrate temperature held constant at 300 °C. An RF power supply was connected to the substrate holder to enable tuning of ion energies, thereby influencing film stress. The metal electrodes were deposited using a 101.6 mm single Pt target in a multi-source chamber. The deposition was carried out at a temperature of 500°C with Ar as the process gas.

The film thickness of AlScN was precisely determined by reflectometry (KLA Filmetrics), with measurements taken at 49 distinct points across the wafer to ensure uniformity assessment. Stress within the films was quantified by measuring wafer bow (see Section S1 of Supplementary Information (SI)), utilizing a Toho stress measurement tool (Toho FLX-2320-R) and applying Stoney's equation[59].

For crystallographic analysis, X-ray diffraction (XRD) in Bragg-Brentano geometry was used to determine the crystallographic orientation and lattice parameters of the AlScN films. The quality of the crystallographic texture was further investigated through X-ray rocking curve measurements. The XRD system used was a Malvern Panalytical X'Pert$^3$ MRD XL diffractometer, equipped with a PIXcel$^3$D detector and a Cu $K_\alpha$ radiation X-ray source (Empyrean Cu LFF HR). The X-ray tube was operated at 40 kV and 40 mA. The incident beam path included a 1/2° fixed incidence slit, and a programmable divergence slit set to 0.03125°, with a W/Si graded parabolic mirror employed for beam focusing. On the diffracted beam side, a 0.16 mm fixed anti-scatter slit was utilized. Surface morphology was examined using Atomic Force Microscopy (AFM) (Park Systems Park NX20), operating in non-contact mode to capture detailed topographical features.

The elemental compositions of the films were thoroughly investigated through a combination of Rutherford Backscattering Spectrometry (RBS) and time-of-flight Elastic Recoil Detection Analysis (ToF-ERDA). Both techniques were performed using the 5 MV NEC-5SDH-2 tandem accelerator at Uppsala University. For RBS measurements, a primary-ion beam of 2 MeV $He^{1+}$ was used, incident at an angle of 5° to the sample-surface normal, and the scattered particles were detected at an angle of 170° relative to the path of the primary beam. The SIMNRA software package[60] was used for the simulation and analysis of the RBS spectra. ToF-ERDA measurements employed a 36 MeV $^{127}I^{8+}$ primary-ion beam at an incidence angle of 67.5° to the sample-surface normal, with recoils detected at an angle of 45° to the path of the primary beam. Raw ToF-ERDA spectra were converted into depth-dependent relative atomic concentration profiles using the Potku software[61]. Cross-sectional scanning transmission electron microscopy (STEM) analyses were carried out using a JEOL JEM-F200, operating at an acceleration voltage of 200 kV and EDX analysis were performed using a dual-X windowless detectors from JEOL with solid angle 1.7sr. The necessary lamellae for TEM analysis were precisely prepared using dual beam Focused Ion Beam system from Thermo Fisher Helios 5 UX and the low energy final cleaning were performed at 5 kV to remove the amorphous layer that degraded the TEM analysis.

For device fabrication, top Pt electrodes were patterned and etched. AZ ECI 3012 resist was coated on the wafers followed by a soft bake at 90 °C for 60s. The wafer was then exposed with an EVG MA 610 mask aligner and post baked for 60s at a temperature of 110 °C. AZ726 MIF developer was used. Post development, a soft $O_2$ plasma was used to clean the sample. Ion beam etching (IBE Veeco Lancer) was used for etching the Pt electrode.

The capacitance-voltage (C-V), leakage current, polarization-electrical loops (P-E) and piezoelectric measurements ($d_{33,f}$) were performed using an aixACCT Systems TF Analyzer 2000 and an aixDBLI double beam laser interferometer, respectively.

### III. RESULTS
#### a. Stress, Microstructure, and Composition

We initiate the deposition process with a stress-free stack of seed layer AlN directly deposited on cleaned Si(100) substrate followed by Pt bottom electrode and $Al_{0.64}Sc_{0.36}N$ film. The AlScN film thickness, its uniformity, and consequently the deposition rate is evaluated. It is determined that the 1σ-uniformity (calculated as the ratio of the standard deviation of the thickness to the mean thickness, multiplied by 100%) is below 1.2%, with a wafer mean $Al_{0.64}Sc_{0.36}N$ film thickness of approximately 500 nm. The deposition rate is estimated to be approximately 8.7 μm/h.

Four distinct films, each with varying biaxial compressive stress ($\sigma_{average}^{bulk}$), were analyzed. The $\sigma_{average}^{bulk}$ values for the films are approximately -100 MPa, -200 MPa, -300 MPa, and -500 MPa, while the stress range ($\sigma_{range}$) across the wafer surfaces are 180 MPa, 230 MPa, 310 MPa, and 475 MPa, respectively. It was evident that an increase in $\sigma_{average}^{bulk}$ corresponded with an increase in $\sigma_{range}$ (see Fig. 1a). For convenience, henceforth the samples are designated as RF100, RF200, RF300, and RF500, based on their respective $\sigma_{average}^{bulk}$ values.

The $\theta$-$2\theta$ scan with Bragg-Brentano geometry revealed diffraction peaks at $2\theta$ values of 36°, 36.2°–36.4°, 40°, and 69° (see Fig. 1b). The peak centered at $2\theta \approx 36°$ is attributed to diffraction from the wurtzite (w) phase AlN-(0002) (underlying seed layer), while the peak centered at $2\theta \approx 36.2°$–$36.4°$ is from (0002) of AlScN. The diffractogram at $2\theta \approx 40°$ is associated with Pt (111), and the Si substrate peak appears at $2\theta \approx 69°$. These results confirm that the films are formed with the w-phase, and no secondary phases are present. The absence of any misoriented planes (other than 0002/c-plane) along the surface of the films further confirms that the films are textured along the c-axis ([0002]) of AlScN. It is worth noting that with an increase in stress, the $2\theta$ of the (0002) diffraction peak shifted to lower $2\theta$ values, and consequently, the interplanar spacing of the (0002) crystal plane increases.

The quality of the texture is further assessed by measuring the XRC-FWHM of the AlScN (0002) (see Fig. 1c). The FWHM of the XRC recorded at the center of the wafer are 1.35°, 1.46°, 1.47°, and 1.59° for the films RF100, RF200, RF300, and RF500, respectively. We also recorded the XRC at the edge

(80 mm from the center) of the wafer. The XRC-FWHM at the edge of the wafer were no more than 0.2° wider than at the center in all cases. With an increase in compressive stress, the XRC-FWHM also increased linearly (see inset of Fig. 1c), leading us to conclude that compressive stress in the films is detrimental to the quality of texture in c-oriented AlScN thin films.

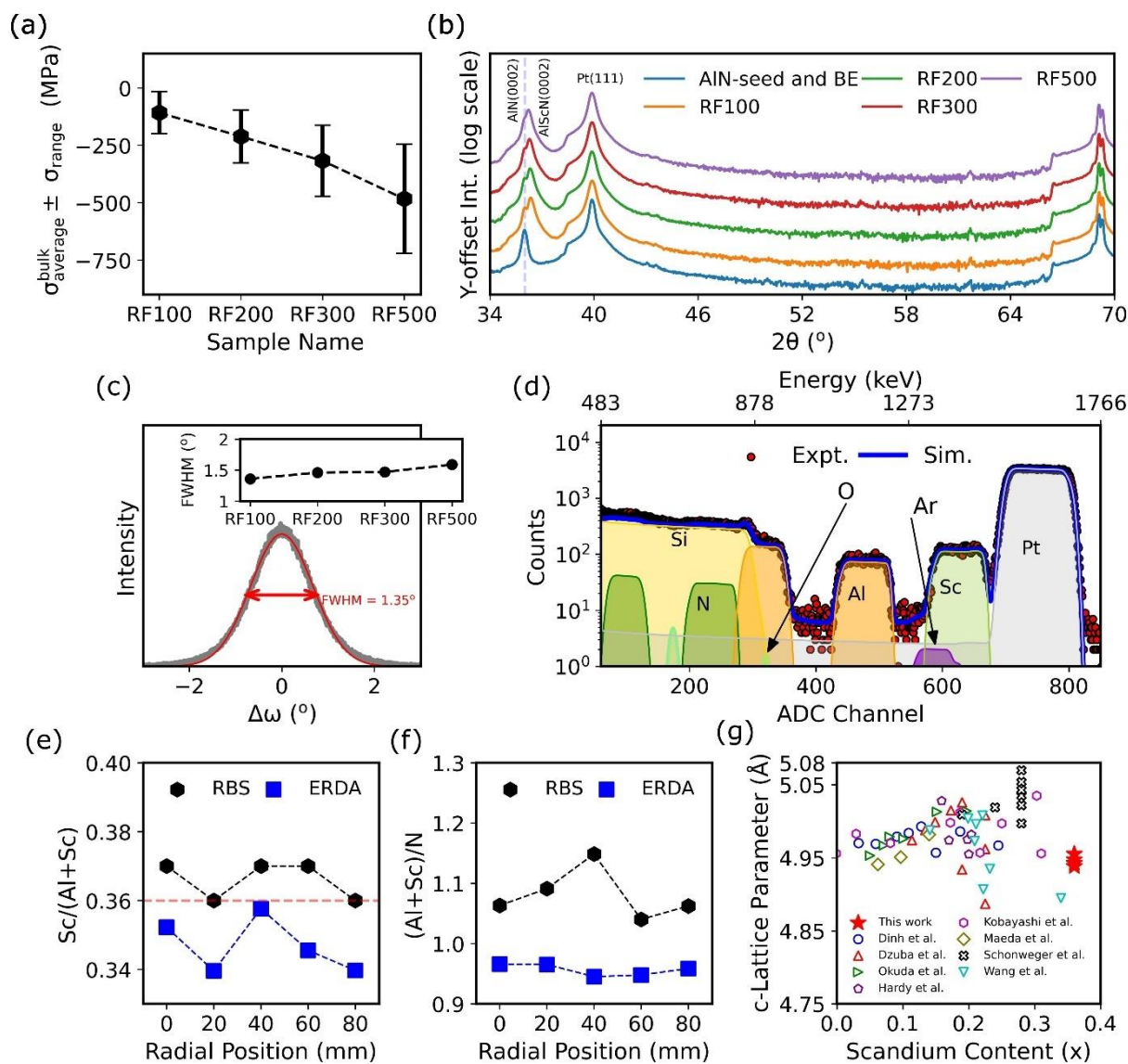

*Figure 1: Figure (a) shows the absolute value of average stress and corresponding stress range for the four investigated samples. Figure (b) presents the θ-2θ XRD patterns of the films, recorded at the wafer center with varying stress, alongside reference diffractograms from the AlN seed layer and Pt bottom electrode. Figure (c) displays the rocking curve of the AlScN (0002) planes for the RF100 sample, with its inset showing the FWHM of the XRC for differently stressed films. Figure (d) provides the RBS spectra of the RF100 film at the wafer center. Figure (e) illustrates the Sc content versus radial position on the RF100 wafer, with the dashed line indicating the Sc content of the magnetron target. Figure (f) details the metal-to-nitrogen ratio in the films as determined by RBS and ERDA. Finally, Figure (g) plots the c-axis parameters against Sc content in the AlScN films, comparing our data with existing literature*

*Dinh et al.[32], Dzuba et al.[62], Okuda et al.[36], Hardy et al.[30], Kobayashi et al.[63], Maeda et al.[64], Schonweger et al.,[35], and Wang et al[65].*

The elemental composition and their depth profiles as well as compositional uniformity over the film surface are determined using a combination of RBS and ERDA techniques. These measurements are performed at five distinct locations of RF100 film: 0 mm (center), 20 mm, 40 mm, 60 mm, and 80 mm (edge) from the wafer's center, as presented in Fig. 1d–f. Fig. 1d depicts the experimentally recorded and simulated RBS spectrum for the RF100 sample obtained at the wafer's center. Clearly the signature of Pt, Al, and Sc is visible. The compositions derived from RBS at these various locations were subsequently compared with those obtained from ERDA. It is evident that both RBS and ERDA exhibit a consistent trend in elemental concentration across the wafer (see Fig. 1e). The Sc atomic fraction ($x = \frac{C_{Sc}}{C_{Al}+C_{Sc}}$, where $C_{Sc}$ and $C_{Al}$ are fraction of Sc and Al in $Al_{1-x}Sc_xN$ alloy) as quantified by RBS at 0 mm, 20 mm, 40 mm, 60 mm, and 80 mm from the wafer's center, was 0.37±0.006, 0.36±0.014, 0.37±0.014, 0.37±0.015, and 0.36±0.014, respectively. The wafer-averaged $x$ is 0.366±0.012, which is marginally higher than the nominal Sc target composition of 0.36. ERDA systematically underestimated the Sc atomic fraction, yielding a wafer-averaged value of 0.347 (see Fig. 1e). The ERDA depth profiles of the constituent elements (see Section S2 of SI) show no significant compositional gradient across the film thickness. The metal-to-nitrogen atomic ratio (Al+Sc)/N was also determined. RBS indicated a slightly metal-rich stoichiometry with a wafer-averaged (Al+Sc)/N is 1.08±0.05, while ERDA estimated this ratio to be 0.956. The observed quantitative discrepancy in elemental compositions between RBS and ERDA is an intrinsic characteristic of these techniques: RBS provides less accurate quantification for lighter elements, both due to the lower scattering cross-sections, and the tendency for signal from these elements to be obscured by those from the substrate. Conversely, ERDA excels at precisely quantifying lighter elements but may be less accurate for heavier elements because of multiple scattering[66,67]. The cumulative concentration of impurities (such as O and C) in the bulk of the AlScN films was consistently below 0.5% of total composition. Therefore, the compositional analysis confirms that the Sc atomic fraction is substantially uniform across the wafer surface and throughout the film thickness, and the wafer-averaged Sc content is in good agreement with the intended target composition.

The average $c$-lattice parameters, obtained from the (0002) and (0004) XRD peaks, are presented in Fig. 1g as a function of $x$ in the $Al_{1-x}Sc_xN$ films. This figure also includes previously reported values from literature for comparison. From existing literature, it is observed that the $c$-lattice parameter typically increases with increasing Sc concentration in $Al_{1-x}Sc_xN$ films until $x$ reaches approximately 0.23 (see Fig. 1g)[26]. Beyond this critical concentration, the $c$-lattice parameter begins to decrease as $x$ increases. Our experimentally estimated $c$-lattice parameters for our films exhibit a similar trend. Specifically, the estimated $c$-lattice parameters for RF100, RF200, RF300, and RF500 are 4.938 Å, 4.942 Å, 4.947 Å, and 4.955 Å, respectively.

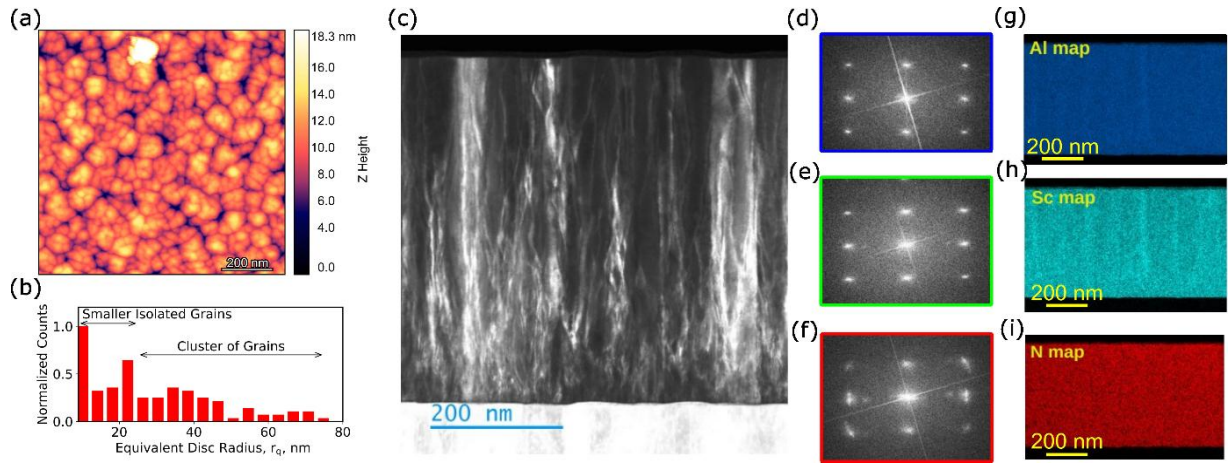

*Figure 2: Characterization of AlScN RF100 thin film microstructure using AFM, TEM, and EDX. (a) AFM reveals the surface morphology at the wafer's center, with (b) detailing the grain size distribution. (c) Dark-field TEM provides insight into the film's internal structure. (d-f) FFT patterns from HRTEM images (top, middle, bottom). (g-i) EDX mapping confirms the elemental distribution of Al, Sc, and N in the film.*

Atomic Force Microscopy (AFM) was employed to characterize the surface morphology and quantify the RMS surface roughness ($S_q$) of the AlScN films. We observed the formation of isolated AOGs, primarily concentrated at the wafer's center, with their density decreasing significantly towards the edges. The maximum surface coverage of AOGs remained below 1% across all scanned areas, indicating high-quality AlScN thin film growth. Notably, AOG density was also sensitive to residual stress within the films, exhibiting a reduction with increasing stress (see Section S3 of SI). The film surface presented a heterogeneous topography, comprising small, isolated grains (predominantly 10-20 nm of equivalent disc radius) alongside larger grain clusters (up to 80 nm radius) (see Fig. 2 a,b). The $S_q$ mapping, provided in the Section S3 of SI, revealed an $S_q$ of 1.59 nm at the center of the RF100 wafer, with values not exceeding 1.70 nm across the entire film. However, an increase in compressive residual stress led to a corresponding increase in $S_q$, reaching up to 2.64 nm at the wafer center and 3.69 nm towards the edge of RF500 wafer.

Dark-field TEM of RF100 film confirmed a uniformly dense microstructure throughout the entire film thickness and in-plane direction, devoid of discontinuous or porous layers (see Fig. 2c). High-resolution TEM images and corresponding Fast Fourier Transform (FFT) patterns are acquired from the bottom, middle, and top regions of the film (Fig. 2 d-f). FFT analysis of the bottom region showed a splitting of in-plane diffraction patterns into two peaks, indicating a slight misorientation of grains during early deposition stages. Conversely, the middle and top regions exhibited a single, sharp FFT peak, signifying a highly oriented film. Energy-dispersive X-ray (EDX) analysis of the AlScN layer confirmed a uniform distribution of Al, Sc, and N throughout the film thickness as well as along the in-plane direction, with no significant compositional gradients. We did not find any evidence of Sc accumulation (see Figure S3 of SI) at the grain boundaries as observed by Wolff et al[68].

b. **Dielectric, Ferroelectric, and Piezoelectric Characterizations**

We characterized the dielectric properties of $Al_{0.64}Sc_{0.36}N$ thin films through comprehensive small-signal capacitance-voltage measurements (CVM), employing a 500 mV AC excitation at 1 kHz. These measurements, performed over an applied electric field range of -1 MV/cm to 1 MV/cm, enabled precise determination of the relative permittivity ($\varepsilon$) and dielectric loss tangent (tan $\delta$), while ensuring operation below the films' polarization switching threshold. In our staircase C–V measurement, a small AC signal (e.g., 500 mV at 1 kHz) is superimposed on each DC voltage step, and the system uses a lock-in method to separate the AC current into in-phase and out-of-phase components. From these, we extracts the conductance G and capacitance C at every bias point and calculates the dielectric loss using the standard relation tan $\delta$ = G / ($\omega$C), where $\omega$ is the angular frequency of the AC signal[69].

Our analyses revealed an average $\varepsilon$ value of approximately 21 across all fabricated wafers, accompanied by an exceptionally low standard deviation of ~ 0.37 (see Fig. 3a). The wafer average $\varepsilon$ of RF200 film deviates from the trend and is 20.6 with a standard deviation of 0.36. The dielectric loss characteristics are quantified through tan $\delta$ measurements (see Fig. 3a). The RF100 sample demonstrated a remarkably low wafer average tan $\delta$ of 0.006, with an associated standard deviation of 0.0005. An increase in compressive residual stress (as shown in Fig. 1a) correlated with an increase in the average tan $\delta$ (e.g., reaching 0.009 with a standard deviation of 0.001 for the high-stress RF500 films, as detailed in the Section S4 of SI). The consistently low tan $\delta$ values observed in the low-stressed films serve as a compelling indicator of the high structural and electrical property of our $Al_{0.64}Sc_{0.36}N$ thin films.

To understand the influence of the residual stress on the electrical properties of our films, we investigated the pre-switching leakage current density ($J_{leak}$) using 500 × 500 μm² square capacitor structures at various location of wafers (see Figure S5(b) and S5(c) of SI). The measurements were performed up to a maximum applied field of ±1 MV/cm, with 0.1 MV/cm step increments and a step duration of 22 seconds. At an applied field of +1 MV/cm, the RF100 film exhibited a wafer average $J_{leak}$ of 0.26 μA/cm² with a standard deviation of 0.05 μA/cm². A crucial observation was the strong correlation between higher stress in films and an increase in both the absolute $J_{leak}$ value and, significantly, the spread of $J_{leak}$ across the wafer (see Fig. 3d). Wafer-scale mapping of $J_{leak}$ further elucidated this trend, demonstrating that capacitors located at the wafer periphery in all films exhibited increased $J_{leak}$ compared to those at the wafer center.

The ferroelectric properties of films are investigated by their current density versus electric field ($J$-$E$) curves and the corresponding polarization versus electric field ($P$–$E$) loops. These measurements are performed by applying triangular pulses at a 5 kHz frequency. A typical box-type $P$-$E$ loop, characteristic of ferroelectric AlScN, is observed (see Fig. 3e) with a slight shift of the $E_C$ toward negative bias, indicating the presence of an internal built-in field. For the RF100 film, the sum of positive and negative remnant polarization (2P$_r$) is approximately 175 μC cm$^{-2}$ with applied maximum field of 4.2 MV/cm. The J–E curve further highlights sharp switching peaks and an asymmetry between positive and negative bias. As the compressive stress increases, $E_C$ also increases, accompanied by greater standard deviation across the wafer (see Fig. 3f and section S5 of SI), indicating non-uniform switching behavior across the wafer, particularly, at the edge of the wafers. For example, the RF500 film has an approximately 20% higher $E_C$ compared to RF100 film to fully switch the switchable capacitors.

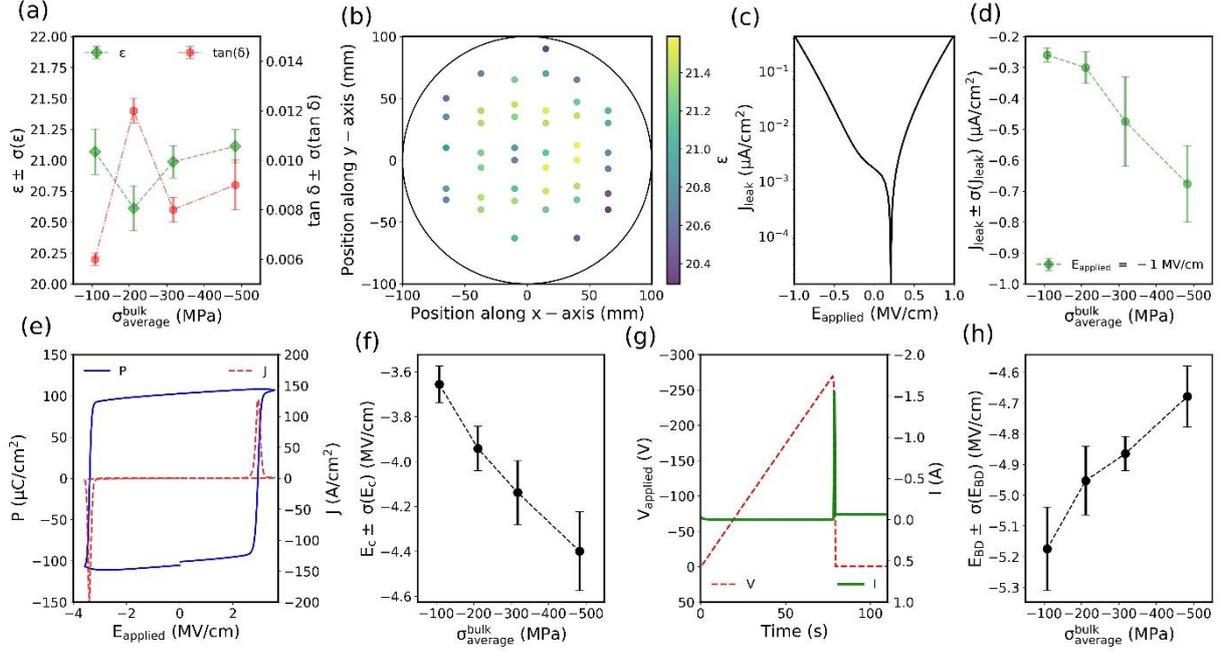

*Figure 3: Electrical characterization of all four wafers investigated in this work. Panel (a) shows wafer average relative permittivity (ε) and loss function (tan δ) of Al$_{0.64}$Sc$_{0.36}$N thin films as a function of compressive stress. Panel (b) shows wafer scale mapping of the ε for RF100 film. Panel (c) shows pre switching leakage current density (J$_{leak}$) vs applied field recorded at the centre of the RF100 film. Panel (d) shows the wafer average J$_{leak}$ vs stress in films. Panel (e) shows P-E and J-E loop recorded at the center of the RF100 film. Panel (f) shows the impact of stress to wafer average coercive field (E$_c$). Panel (g) shows proto typical curves of applied voltage vs time and measured current vs time in breakdown test measurements. Wafer average breakdown voltages vs stress is presented in panel (h).*

The dielectric breakdown measurement is carried out with a maximum voltage of 400 V, corresponding to an electric field of 8 MV/cm. A voltage step of 2 V with a step duration of 500 ms is applied on a 500×500 μm$^2$ capacitor. Measurements were carried out in the negative field direction, as the negative side exhibited higher leakage and was therefore assumed to be electrically weaker. At the center of the RF100 wafer (see Fig. 3g), breakdown occurred at an applied voltage of approximately –260 V, corresponding to an electric field of about –5.2 MV/cm. Interestingly, in contrast to $E_C$, as the $\sigma_{average}^{bulk}$ values increase from the 100 MPa (RF100), the average breakdown fields ($E_{BD}$) decreased (see Fig. 3h), resulting in a narrow switching field window. For example, RF100 exhibited an $E_{BD}$:$E_C$ ratio of around 1.42, whereas the highly stressed RF500 film showed a reduced ratio of 1.06. This indicates that higher $\sigma_{average}^{bulk}$ makes the film more prone to electrical breakdown during switching and is also associated with an increased switching failure rate.

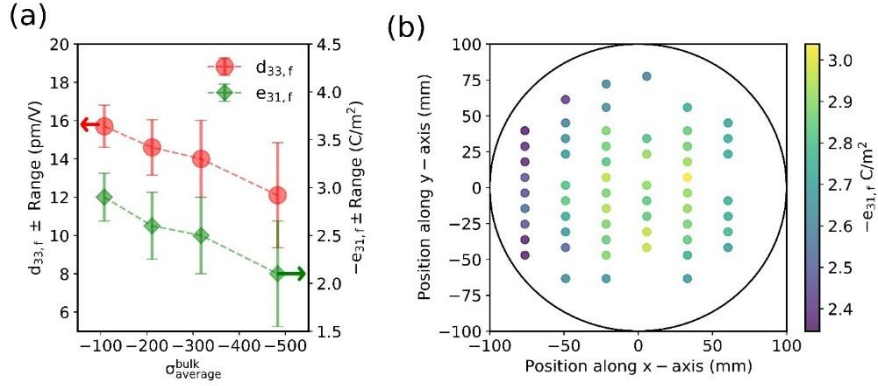

Figure 4: Piezoresponse of $Al_{0.64}Sc_{0.36}N$ thin films. Figure (a) shows the $d_{33,f}$ and $e_{31,f}$ vs residual stress in the film. Figure (b) shows the wafer scale mapping of the $e_{31,f}$ of RF100 film.

Piezoelectric properties are comprehensively evaluated, focusing on the effective longitudinal piezoelectric strain coefficient, $d_{33,f}$, and the transverse effective piezoelectric stress coefficient, $e_{31,f}$. The electrode-size dependence of $d_{33,f}$ of piezoelectric thin films measured by double beam laser interferometry as proposed by Sivaramakrishnan et al[70]. Four different electrode per die with a with diameters of 600, 800, 1000, and 1200 μm is used.

True $d_{33,f}$ and $e_{31,f}$ values are obtained by a linear fitting, involving all the pad size dependence of measured $d_{33,f}$ values ($d_{33,f,meas}$) as given by eq.1.

$$d_{33,f,meas}(r) = d_{33,f} - f(r) \cdot \frac{v_s}{Y_s} \cdot e_{31,f}$$

Where $f(r)$ describes the change in the relative magnitude of in-plane and out of plane stresses as a function of electrode size, $r$, and are often determined from FEM simulations. $v_s$ and $Y_s$ are mechanical parameters Poisson's ratio and Youngs modulus of the Si substrate. In our work, we used program from aixACCT Systems. The technical details of the implementation are presented elsewhere[70].

For the RF100, the wafer median $d_{33,f}$ is measured at 15.7 pm/V and exhibiting a range of ±2.2 pm/V across the wafer (see Fig. 4 a). We observed that increasing compressive $\sigma_{average}^{bulk}$ led to a reduction in the median $d_{33,f}$ and wider range, with the median value dropping to 12.1 pm/V and the range expanding to ±5.5 pm/V at ≈ -500 MPa of stress (see Fig. 4a). The RF100 wafer also showed a median $e_{31,f}$ of -2.90 C/m² with a range of ±0.5 C/m². A similar trend is observed for $e_{31,f}$ where its magnitude decreased with increasing compressive stress of films. The RF500 sample displayed a median $e_{31,f}$ of -2.1 C/m² with a range of ±1.1 C/m². Wafer-scale mapping of $e_{31,f}$ for RF100 (see Fig. 4b) revealed a distinct donut-shaped profile, with slightly lower values at the wafer edges compared to the center, while maintaining good overall uniformity. Similar trends were consistently observed across other wafers. To our knowledge, these reported $d_{33,f}$ and $e_{31,f}$ values are among the highest documented for $Al_{1-x}Sc_xN$ in the literature for films grown with industrial scale production tool and on large-area Si wafers. The 1σ uniformity of the $d_{33,f}$ and $e_{31,f}$ data as observed in RF100, RF200, RF300 and RF500 wafers are 2.9%, 2,7%, 4.3%, and 6.9%, respectively.

## IV. DISCUSSION

The observed increase in biaxial compressive stress with increasing RF bias power is attributed to the "atomic peening" effect. The seminal work by Chason et al. [71] provides a kinetic model for stress formation and evolution in energetic vapor processes like magnetron sputtering. In polycrystalline thin films, the formation of compressive stress is linked to two main mechanisms: grain boundary densification ($\sigma_{gb}$) and the formation of point defects within the bulk of the grains ($\sigma_{bulk}$). The $\sigma_{gb}$ component is directly proportional to the length of the dense grain boundary and inversely proportional to the grain sizes (eq. 2), while the $\sigma_{bulk}$ component is directly proportional to the number of defects and their depth of formation (eq.3) in the bulk of grains.

$$\sigma_{gb} \propto l/d \qquad (-2)$$

$$\sigma_{bulk} \propto \left(1 - \frac{l}{d}\right) \qquad (-3)$$

With increasing RF bias power, incoming ions gain additional kinetic energy and bombard the growth front, leading to the formation of point defects in the bulk of the grains. Due to their low mobility and their formation far from the film's growth surface, these defects are often trapped in the bulk, contributing to the compressive stress. As the RF bias power increases, both the depth at which point defects form and their number density increases, resulting in a corresponding increase in compressive stress. High bombardment of ions on the thin film growth front also led to increase in the roughness and consequently increase in the mosaicity of newly nucleated grains during film deposition.

A statistical Pearson's correlation[72] between microstructure and electrical properties is evaluated and its correlation coefficient (p) matrix is presented as Figure 5a for these samples. Evidently, a weaker correlation between microstructure such as stress and its range, and FWHM of the rocking curve to ε is seen, hinting that ε is largely related to composition rather than film's microstructure. The ε for $Al_{0.64}Sc_{0.36}N$ is slightly larger than $Al_{0.70}Sc_{0.30}N$ (ε = 18.5 at frequency = 1 kHz) thin films, as obtained with a similar measurement with a comparable stress in the films[53]. The tan δ shows an increasing trend with higher compressive stress. The increase in the tan δ and leakage current with compressive stress is largely associated with the formation of point defects induced trap states due to bombardment of high energetics ions during the film growth.

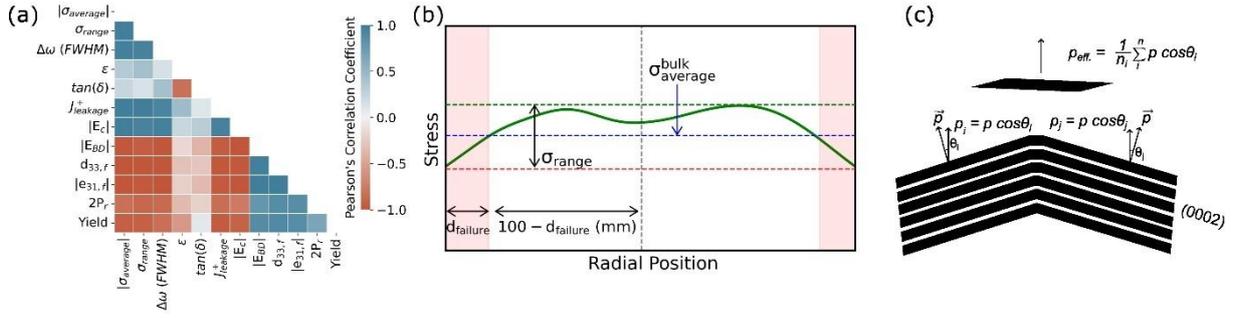

*Figure 5: (a) shows the correlation matrix plot of microstructure and electrical properties of $Al_{0.64}Sc_{0.36}N$ films. Panel (b) shows the stress profile of films across the diameter of the wafer. The $d_{failure}$ is the region where ferroelectric capacitors failed to switch. The remaining (100- $d_{failure}$) mm is the radius of the circle within which capacitors are fully switched. Panel (c) shows the impact of tilt of grains to the average dipole moment per unit area.*

The high correlation (p > .99) between $\sigma_{average}^{bulk}$ and $J_{leak}$ indicates that compressive stress in AlScN films adversely influences the leakage current. This is consistent with the previous reports in the literature [73–75], where leakage current in a metal/AlScN/metal capacitor is attributed to trap-assisted Poole–Frenkel (P-F) emission. The trap states are speculated to originate from nitrogen vacancies ($V_N$) that are formed in the bulk of AlScN film and at the metal–AlScN interfaces. These point defects induced trap states and their accumulation near the metal/AlScN interfaces reduces the Schottky barrier height, facilitating electrical conduction in the form of leakage current. Our P-F fitting (ln(J/E) vs $E^{0.5}$) showed the linear relationship only at higher E values and a deviation from the linear behaviour at lower E (see Section S5 of SI), indicating that at high bias the leakage current is dominated by P-F mechanism. In the low bias region different mechanisms e.g., Schottky emission, tunnelling, hopping mechanism may involves [76]. A detailed analysis is needed to clearly establish the dominant mechanism and is currently beyond the scope of the present paper. Increase in $J_{leak}$ with compressive stress in films can be understood as follows:

As the RF bias power increases the $V_N$ and/or associated defects concentrations increases and consequently the carrier trap densities in the electronic structure of bulk of AlScN thin films. These trap states facilitate bulk limited leakage conduction. Formation of these point defects generate residual stress in the films. We note that in samples and regions where the residual stress is higher $S_q$ is also higher. With higher $S_q$ of AlScN films, the interfacial DOS of AlScN and top Pt electrode is higher. Interfacial DOS can also effectively reduce Schottky barrier height by downward band bending and promote leakage conduction[77,78]. Our investigation suggests that the leakage current with high bias (E > 0.4 MV/cm) in these samples is largely dominated by bulk limited conduction Poole-Frenkel mechanism (see Section S5 of SI).

The experimentally observed increase in $|E_C|$ with compressive stress in AlScN thin films can be explained by the formation of dipolar defects [79]. Evidence for the formation of the dipolar defects is seen from the wafer-averaged $E_{C+}$: $E_{C-}$ ratio, which remains slightly below unity and decreases further with rising compressive stress (see Section S7 of SI). Formation of these dipolar defects promote "dipole pinning" that increases $|E_C|$.

*Table 1: Summary of process-structure-properties of AlScN deposited with different techniques.*

| Properties | This work | Kreutzer et al.[56] | Casamento et al.[80] | Wingqvist et al.[81] | Lin et al.[82] | Moreira et al.[83] |
|---|---|---|---|---|---|---|
| Composition | $Al_{0.64}Sc_{0.36}N$ | $Al_{0.6}Sc_{0.4}N$ | $Al_{0.75}Sc_{0.25}N$ | $Al_{0.7}Sc_{0.3}N$ | $Al_{0.7}Sc_{0.3}N$ | $Al_{0.85}Sc_{0.15}N$ |
| Deposition Methods | Sputtering with alloy target $Al_{0.64}Sc_{0.36}$ | Sputtering with alloy target $Al_{0.6}Sc_{0.4}$ | MBE | Co-sputtering | Sputtering with alloy $Al_{0.7}Sc_{0.3}$ target | Co-sputtering |
| Template | Pt/AlN/Si(100) | Pt/Ti/*a*-SiO$_2$/Si(100) | GaN bulk substrate | TiN(111)/Sapphire(0001) | Mo/*a*-SiO$_2$/Si(100) | Mo/AlN/Si(100) |
| Orientation | 0001-textured | 0001-textured | Epitaxial AlScN | Epitaxial AlScN | 0001-textured | 0001-textured |
| Film Thickness (nm) | 500 | 595 | 100 | 250-500 | 100-1000 | 1200 |
| XRC-FWHM (0002) (°) | 1.35-1.59 | 1.40-5.01 | - | - | 2.00-1.30 | 2.00 |
| $S_q$ (nm) | 1.59-1.70 | - | - | - | 0.8-1.1 | - |
| $\varepsilon_r$ | 20.6 at $f$ = 1 kHz | 28.3 at $f$ = 5 kHz | 22 at $f$ = 500 kHz | 13.8 | - | 14.1 |
| tan δ (%) | 0.6 at $f$ = 1 kHz | 0.57 at $f$ = 5 kHz | 0.016 at $f$ = 500 kHz | 0.8 | 0.5-0.2 | - |
| $J_{leak}$ (μA/cm$^2$) | 0.26 (E = 1MV/cm) | - | - | - | - | - |
| $d_{33,f}$ (pm/V) | 15.7 | 15.62 | - | - | 9.5 to 11.3 | - |
| $e_{31,f}$ (C/m$^2$) | -2.9 | - | - | - | -1.76 to -2.09 | - |

It is quite commonly observed in the literature [20,84,85] that the $d_{33,f}$ and/or $e_{31,f}$ of $Al_{1-x}Sc_xN$ is better when XRC-FWHM of 0002 plane is < 2°. In our work, we observe that the $d_{33,f}$ and $e_{31,f}$ has a very strong correlation (p ≈ .99) with the XRC-FWHM of 0002 planes along with bulk stress value (see Fig. 5a). In polycrystalline thin films, XRC-FWHM of the planes perpendicular to growth direction characterizes the quality of grains orientation along the growth direction and lower width indicates lower tilt. With higher XRC-FWHM the number of grains with perfect c-axis orientation also reduces. In Figure 5c we depict the role of these misoriented grains to the net effective dipole moment per unit area ($p_{i,j}$) compared to ideal (when no misorientation occur) dipole moment (p) per unit area. It is quite clear that with higher XRC-FWHM the angles between the normal to film-substrate and the c-axis ($\theta_{i,j}$) of grown grains are larger. With increase in $\theta_{i,j}$, the net effective $p_{i,j}$ will be normalized by a factor $\cos\theta_{i,j}$. Our observation of reduction in $d_{33,f}$ with wider XRC-FWHM agreed well with this hypothesis. Along with XRC-FWHM, $d_{33,f}$ and $e_{31,f}$ also shows high correlation with bulk stress value which indicate that the point defect induced stress is impacting the $d_{33,f}$ and $e_{31,f}$. A good 1σ uniformity in $d_{33,f}/e_{31,f}$ that is important for industrial adoption is achieved in low stressed film (RF100 and RF200) and is attributed to high quality c-axis texture and its uniformity of as well as low compressive stress and lower stress range across the diameter of wafers.

The stress dependence of electrical properties ε, tan δ, $J_{leak}$, $E_c$, $E_{BD}$, $d_{33,f}/e_{31,f}$ is similar to the dipolar defect concentration vs electrical properties trend [79,86]. This suggests that with increase in RF bias power during film growth, dipolar defects are forming and are also source of compressive stress in films. Our observation suggest that formation of higher concentration of these dipolar defects is detrimental to the piezo- and ferroelectric properties of the AlScN. A detailed comparison in structure-properties is presented in Table 1 along with values previously reported in the literature. Clearly, the microstructural and piezoelectric properties of our films are superior to others while relative permittivity and dielectric loss values are similar to sputtered thin films reported in the literature.

To this end, we demonstrate the development of high-quality wafer scale thin film of $Al_{0.64}Sc_{0.36}N$ using a 300 mm alloy target in an industry compatible deposition tool. Through process optimization and appropriate selection of seed layers and electrode we achieve superior microstructural quality and wafer uniformity that resulted into high electromechanical response of the $Al_{0.64}Sc_{0.36}N$ thin films. Excessive compressive stress in AlScN films degrades their piezoelectric and ferroelectric properties and reduces wafer yield, thus we suggest manufacturers to minimize film's compressive stress for optimal performance. Importantly, we establish the reproducibility of our method. Repeating the process in two separate tools with three different targets of varying lifetimes yielded both qualitatively similar and quantitatively repeatable results. This robustness and transferability confirm our technique's direct applicability in semiconductor foundries, enabling reliable, and scalable production.

## V. SUMMARY

We developed high-performance piezoelectric and ferroelectric $Al_{0.64}Sc_{0.36}N$ thin films on 200 mm silicon wafers and thoroughly investigated how residual compressive stress influences their microstructure and electromechanical properties. Using a commercial tool with an alloy target, we deposited films that consistently exhibit a strong 0002 orientation, regardless of stress levels, demonstrating robust crystallographic texture. We confirmed the target composition through RBS and ERDA, achieving excellent uniformity across the wafer without elemental accumulation, such as Sc segregation. Through X-ray rocking curve analysis of the 0002 planes, we measured a FWHM of 1.35° for low-stress films, which increases to 1.59° at 500 MPa of compressive stress, ranking among the best reported for polycrystalline textured films. Our microstructural studies reveal dense films where smaller grains bundle into larger grains, contributing to superior structural quality. This quality manifests itself in the electrical properties: low-stress films achieve a wafer-average relative permittivity, ε, of 21 and an exceptionally low tangent loss, tan δ, of 0.006. We recorded a leakage current of 0.26 μA/cm² at 1 MV/cm in low-stress films, which rises with increasing stress. Our ferroelectric characterization shows that higher compressive stress increases coercive fields and reduces breakdown fields, with devices near wafer edges failing at bulk stresses above 300 MPa, thus lowering the yield. In low-stress films, we measure outstanding piezoelectric properties, including a longitudinal piezoelectric coefficient, $d_{33,f}$, of 15.7 pm/V and a transverse effective piezoelectric stress coefficient, $e_{31,f}$ of -2.90 C/m², among the highest reported in an industrial scale process technologies and on large area substrates. These properties decline as compressive stress increases. We attribute the stress-dependent behaviour to dipolar defects formed during deposition, driven by increased RF bias power that boosts ion kinetic energy and induces defects in the bulk and at the growth front, resulting in residual compressive stress. The exact atomic configuration of dipolar defects needs further investigation. Minimizing stress is essential for high wafer yield and optimal electromechanical performance. Our process and materials align with high-throughput semiconductor foundry production, positioning $Al_{0.64}Sc_{0.36}N$ thin films as a promising candidate for advanced piezoelectric and ferroelectric applications.

## SUPPLEMENTARY MATERIAL

The supplementary information contains the method used for the determination of local stress, depth profile of ERDA data, full AFM scans across the diameter of the wafers, wafer-scale mapping results of electrical properties, and the mathematical fitting of electrical data.

## AUTHOR DECLARATIONS

**Conflict of Interest**

The authors have no conflicts to disclose

**Authors Contribution:**

**SN**: Thin Film Development, Materials Characterizations, Data Analysis, Preparation of Manuscript. **VRN**: Electrical Characterization and Failure Analysis, Wafer Level Mapping, Data Analysis, Manuscript Editing. **RSB**: Piezoelectric Characterization, Data Analysis, Manuscript Editing. **DS**: Material Characterizations and Data Analysis. **DH**: Thin Film Development, and Materials Characterizations. **WRA**: Microfabrication. **CML**: Materials Characterizations, Data Analysis. **RJWF**: Experiments and Compositional Data Analysis of IBA. **AS and DC**: TEM Sample Preparation and Measurements. **AB**: Electrical Characterization and Data Validation. **RS**: Electrical Characterization and Data Validation. **EA**: Electrical Characterization and Data Validation. **AG**: Electrical Data Analysis and Validation. **MK**: Project Management and Thin Film Data Validation. **AP**: Electrical Characterization and Failure Analysis, Wafer Level Mapping, Data Analysis and Data Validation. **SC**: Project Management, Data Analysis and Data Validation. **AR**: Project Management and Device Conceptualization. **MM**: Project Management and Conceptualization. **CH**: Project Management and Data Validation. **MD**: Project Management and Data Validation. **ADP**: Project Management, Project Conceptualization, Device Design, Microfabrication, Data Validation.

## DATA AVAILABILITY

The data that support the findings of this study are available from the corresponding author upon reasonable request.